\documentclass[floats,floatfix,showpacs,amssymb,prd,superscriptaddress,nofootinbib,twocolumn,aps,eqsecnum]{revtex4-1}
\pdfoutput=1

\usepackage{amsmath, amsthm, amsfonts, amssymb, latexsym}
\usepackage{graphicx}
\usepackage{url}

\usepackage{cprotect}

\usepackage[unicode]{hyperref}
\hypersetup{
    unicode=true,          
    pdftitle={Collisions of oppositely charged black holes},
    pdfsubject={Numerical Relativity},
    pdfcreator={LaTeX},          
    pdfproducer={LaTeX},         
    pdfkeywords={General Relativity, Numerical Relativity, Black Holes},
    colorlinks=false,            
  }

\begin{document}

\title{Collisions of oppositely charged black holes}

  \author{Miguel~Zilh\~ao}\email{mzilhao@astro.rit.edu}
  \affiliation{
    Center for Computational Relativity and Gravitation and School of Mathematical Sciences, 
    Rochester Institute of Technology, 
    Rochester, NY 14623, USA
  }

  \author{Vitor~Cardoso}
  \affiliation{
    CENTRA, Departamento de F\'{\i}sica, Instituto Superior T\'ecnico, Universidade de Lisboa,
    Avenida Rovisco Pais 1, 1049 Lisboa, Portugal.
  }
  \affiliation{
    Perimeter Institute for Theoretical Physics, Waterloo, Ontario N2L 2Y5, Canada
  }

  \author{Carlos~Herdeiro}
  \affiliation{
    Departamento de F\'\i sica da Universidade de Aveiro and I3N, 
    Campus de Santiago, 3810-183 Aveiro, Portugal
  }
  
  \author{Luis~Lehner}
  \affiliation{
    Perimeter Institute for Theoretical Physics, Waterloo, Ontario N2L 2Y5, Canada
  }

  \author{Ulrich~Sperhake}
  \affiliation{
    Department of Applied Mathematics and Theoretical Physics,
    Centre for Mathematical Sciences, University of Cambridge,
    Cambridge CB3 0WA, UK
  }
  \affiliation{
    Theoretical Astrophysics 350-17,
    California Institute of Technology,
    Pasadena, CA 91125, USA
  }
  \affiliation{
    Department of Physics and Astronomy,
    The University of Mississippi,
    University, MS 38677, USA
  }

\date{November 2013} 

\begin{abstract}
  The first fully non-linear numerical simulations of colliding charged black
  holes in $D=4$ Einstein-Maxwell theory were recently
  reported~\cite{Zilhao:2012gp}. These collisions were performed for black holes
  with equal charge-to-mass ratio, for which initial data can be found in closed
  analytic form. Here we generalize the study of collisions of charged black
  holes to the case of unequal charge-to-mass ratios. We focus on oppositely
  charged black holes, as to maximize acceleration-dependent effects. As $|Q|/M$
  increases from 0 to 0.99, we observe that the gravitational radiation emitted
  increases by a factor of $\sim 2.7$; the electromagnetic radiation emission
  becomes dominant for $|Q|/M \gtrsim 0.37$ and at $|Q|/M=0.99$ is larger, by a
  factor of $\sim 5.8$, than its gravitational counterpart. We observe that
  these numerical results exhibit a precise and simple scaling with the
  charge. Furthermore, we show that the results from the numerical simulations
  are qualitatively captured by a simple analytic model that computes the
  electromagnetic dipolar radiation and the gravitational quadrupolar radiation
  of two non-relativistic interacting particles in Minkowski spacetime.
\end{abstract}


\maketitle


\section{Introduction}
Astrophysical black hole (BH) collisions are expected to release
tremendous amounts of energy through gravitational waves. For instance, a binary system of two
non-spinning equal mass BHs is expected to release over 3\% of its
total energy into gravitational radiation, during the plunge/merger
phase~\cite{Pretorius:2007nq}. This amounts to a peak luminosity of about
$10^{56}~{\rm erg/s}$, i.e.~$\sim 10^{23}~L_{\odot}$.
One may then ask: how large can
the efficiency of any BH collision be, in converting the energy of the
BHs into radiation?

The simplest argument that bounds this efficiency was provided by
Hawking and is based on the area theorem~\cite{Hawking:1971}. The
second law of thermodynamics indicates that for the (head-on) collision
of two Schwarzschild BHs starting from rest, no more than 29\% of the
initial energy contained in the BHs can be converted into gravitational
radiation. For BHs colliding head-on at very high energies, on the
other hand, a completely different argument originally due to Penrose
\cite{Penrose1974}, based on the existence of an apparent horizon in
the head-on collision of shock waves, curiously yields exactly the
same bound of 29\% (see \cite{Eardley:2002re} for its $D$-dimensional
generalization). 
The former estimate turns out to be an
extremely conservative limit, as numerical relativity simulations of BH collisions show that
for BHs starting from rest the energy released is much lower---of the order of 0.05\%
of the total energy~\cite{Witek:2010xi}. On the other hand, the latter estimate
is only off by a factor of $\simeq 2$, as the energy released
for high energy collisions approaches 14\% in the ultrarelativistic
limit \cite{Sperhake:2008ga}, a mismatch by merely a factor of 2 with the shock
waves estimate.

The efficiency in converting the energy of a system of BHs into radiation
can be further changed by introducing impact parameter and spin.  Collisions from rest of BHs with aligned spins perpendicular to the
collision axis radiate $0.118\%$ of the total energy when $a/M \simeq
0.4$~\cite{Campanelli:2006fg}, while for the anti-aligned case
a lower radiation efficiency of $0.090\%$ has been found
for $a/M \simeq 0.75$~\cite{Choi:2007eu}.
For BH binaries in quasicircular orbits, spins aligned with the
orbital angular momentum significantly increase the amount of
energy radiated in gravitational waves~\cite{Campanelli:2006uy}. Extrapolated to
the extremal limit $a/M=1$, the prediction is $E_{\rm rad}/M=11.40\%$~\cite{Hemberger:2013hsa}.
A more recent analysis~\cite{Sperhake:2012me} has further led to the
following two conclusions: (i) for high energy collisions of spinning
BHs with impact parameter, in the ultra-relativistic limit,
about 50\% of the energy can be radiated away; (ii) for $v/c\gtrsim 0.9$
spin effects become washed away. This latter observation serves to
support the idea that \textit{matter does not matter} at very high
energies, since processes should be dominated by the kinetic energy
and hence details of the internal structure of the colliding
objects, as, for example, spin or charge,
should become irrelevant. Earlier evidence for this hypothesis had
already been provided by high energy collisions of boson fields
and fluid particles~\cite{Choptuik:2009ww,Rezzolla:2012nr,East:2012mb}.
The former observation, on the other hand, finds
a curious parallel in higher dimensional head-on collisions of shock
waves: both, apparent horizon arguments \cite{Eardley:2002re} and
perturbative analysis \cite{Herdeiro:2011ck,Coelho:2012sy}, suggest
that the radiative efficiency is always smaller than 50\%
approaching this value in the limit of an infinite number of dimensions.

The profound influence that the additional physical parameters discussed
above (spin and impact parameter) have in the total radiated energy
prompts the question of whether charges can also affect the outcome. 
Do collisions of charged BHs support these two observations? That is,
can no more than 50\% of the energy be radiated away in any BH collision
and the details of the BH structure---namely its charge---become
irrelevant at very high energies? In particular, concerning the first
point, Hawking's area theorem argument suggests that the analysis of
charged BHs may be of special relevance, as a simple comparison with
the rotating case reveals. In a head-on collision of equal mass $M$,
anti-aligned spins $\pm aM$ Kerr BHs\footnote{To have a head-on collision of Kerr BHs starting from rest, their spins must be either
aligned or anti-aligned; to maximize accelerations and hence the emitted
radiation we choose anti-aligned spins, since the spin-spin force becomes
attractive~\cite{Wald:1972sz}.}, starting from rest, the fraction of
radiated energy is bounded by
\[ \epsilon\le 1-\frac{1}{2}\sqrt{1+\sqrt{1-(a/M)^2}} \ ,\]
which varies from 29\% to 50\% as $|a|$ varies from zero to $M$. On the
other hand, in a head-on collision of equal mass $M$, opposite charge
$\pm Q$ Reissner-Nordstr\"om BHs, starting from rest, the fraction of
radiated energy is bounded by
\[ \epsilon\le 1-\frac{1}{2\sqrt{2}}\left(1+\sqrt{1-(Q/M)^2}\right) \ ,\]
which varies from 29\% to 65\% as $|Q|$ varies from zero to $M$.   In spite
of this argument providing a poor estimate of the
actual value, it indicates a larger increase in $E_{\rm rad}$
if the maximum amount
of charge is added to each BH in comparison with adding instead
the maximum spin. High energy collisions of oppositely charged BHs
provide particularly interesting problems to test the
aforementioned observations.

The first step towards this goal was taken in~\cite{Zilhao:2012gp}, hereafter
referred to as Paper~I, where collisions of charged BHs in
Einstein-Maxwell theory were studied. That work was restricted to the case of
head-on collisions from rest of BHs with equal charge-to-mass ratios, which
admits analytic initial data.  In this paper, an important extension is given as
we study collisions of charged BHs with unequal charge-to-mass ratio;
in particular we consider oppositely charged BHs in view of the above
motivations. These configurations require a numerical construction of initial
data. We generate such initial data using a modified version of the
\textsc{TwoPunctures} spectral solver~\cite{Ansorg:2004ds} that is described in
Sec.~\ref{sec:evol-eq} along with a summary of the formulation of the
Einstein-Maxwell equations used in our code. In
Sec.~\ref{classical_expectations} we produce analytic estimates based on
considerations of point charges and masses in flat space using electromagnetic
dipolar emission and quadrupolar gravitational emission approximations. We show
that these estimates can reproduce qualitatively---and for an appropriate value
of a cut-off parameter even quantitatively---the results of the numerical
simulations. The fully non-linear numerical simulations in Einstein-Maxwell
theory are presented in Sec.~\ref{sec:numerical_results} where, in particular, a
simple and precise scaling of the waveforms and of the radiated energy with the
charge is observed. Concluding remarks are presented in Sec.~\ref{sec:final}.

\section{Formalism}
\label{sec:evol-eq}
As in Paper~I, we shall consider the enlarged electro-vacuum
Einstein-Maxwell equations:
\begin{equation}
  \label{eq:EFE}
  \begin{aligned}
    R_{\mu \nu} - \frac{R}{2} g_{\mu \nu} & = 8\pi T_{\mu \nu} \  ,\\
    \nabla_{\mu}\left( F^{\mu \nu} + g^{\mu\nu} \Psi
    \right) & = -\kappa n^{\nu} \Psi \ , \\
    \nabla_{\mu}\left(
    \star \!{}F^{\mu \nu} + g^{\mu\nu} \Phi
    \right) & = -\kappa n^{\nu} \Phi \ ,
  \end{aligned}
\end{equation}
where $F^{\mu \nu}$ is the Maxwell tensor and $\star \!{}F^{\mu \nu}$ its Hodge dual, $\kappa$ is a constant and
$n^\mu$ is the 4-velocity of the Eulerian observer.  We recover the
standard Einstein-Maxwell system when $\Psi = 0 = \Phi$.
With the scalar field $\Psi$ and pseudo-scalar $\Phi$ introduced in
this way, the evolution of this system drives $\Psi$ and $\Phi$
to zero (for positive $\kappa$), thus ensuring the magnetic and electric
constraints are controlled~\cite{Komissarov:2007wk,Palenzuela:2008sf}.
The electromagnetic stress-energy tensor takes the usual form
\begin{equation}
  \label{eq:Tmunu}
  T_{\mu \nu} = \frac{1}{4\pi} \left[ F_{\mu}{}^{\lambda} F_{\nu \lambda}
    - \frac{1}{4} g_{\mu \nu} F^{\lambda \sigma} F_{\lambda \sigma}
    \right] \ .
\end{equation}

The 3+1 decomposition was detailed in Paper~I. Here we recall that we introduce a
3-metric
\begin{equation}
  \gamma_{\mu\nu} = g_{\mu \nu} + n_{\mu} n_{\nu} \ ,  \label{eq:3metric}
\end{equation}
and we denote by $D_i$ the covariant derivative associated with $\gamma_{ij}$, where $i,j=1,2,3$ are spatial indices. The Maxwell tensor and its dual are decomposed in the electric and magnetic 4-vectors as
\begin{equation}
  \begin{aligned}
    F_{\mu \nu} & = n_{\mu} E_{\nu} - n_{\nu} E_{\mu}
      + \epsilon_{\mu\nu\alpha\beta} B^{\alpha} n^{\beta}  \ ,\\
      \star \! F_{\mu \nu} & = n_{\mu} B_{\nu} - n_{\nu} B_{\mu}
      - \epsilon_{\mu\nu\alpha\beta} E^{\alpha} n^{\beta}  \ ,
  \end{aligned}
  \label{eq:faraday}
\end{equation}
where we use the convention $\epsilon_{1230} = \sqrt{-g}$,
$\epsilon_{\alpha \beta \gamma}
= \epsilon_{\alpha \beta \gamma \delta} n^{\delta}$,
$\epsilon_{123} = \sqrt{\gamma}$.

In Paper~I, BH binaries with equal charge and mass
colliding from rest were considered.  Such configurations allow for initial
data to be specified in fully analytical form using the Brill-Lindquist
construction~\cite{Brill:1963yv}. We here want to consider BH binaries
with different charge-to-mass ratios, which no longer admit this simple construction. We will thus follow the procedure presented
in~\cite{Alcubierre:2009ij}, which we outline in the following.

Assuming time-symmetric initial configurations, i.e. such that the extrinsic curvature vanishes, $K_{ij}=0$, combined with the condition of an initially vanishing
magnetic field,
the magnetic constraint $D_i B^i=0$ and momentum constraint
are automatically satisfied. By further assuming the spatial metric
to be conformally flat
\begin{equation}
  \gamma_{ij} dx^i dx^j = \psi^4 \left( dx^2 + dy^2 + dz^2 \right) \ ,
  \label{eq:inigamma}
\end{equation}
the Hamiltonian constraint reduces to
\begin{equation}
  \label{eq:ham-0}
  \triangle \psi + \frac{1}{4} \psi^9 E^i E^j \delta_{ij} \ = 0 ,
\end{equation}
where $\triangle$ is the flat space Laplace operator. The electric
constraint, Gauss's law, has the usual form
\begin{equation}
\label{eq:elec-constraint}
  D_i E^{i} = 0 
\end{equation}
and can be solved independently of~\eqref{eq:ham-0}. Introducing an electric potential $\varphi$ through
\begin{equation}
\label{eq:elec-potential}
E^i = - \psi^{-6} \delta^{ij} \partial_j \varphi \,,
\end{equation}
we find that
\begin{equation}
\label{eq:varphi}
    \varphi = \sum_{i=1}^{N}
        \frac{ q_i }{|\mathbf{x} - \mathbf{x}_i|} \,,
\end{equation}
where $\mathbf{x}_i \equiv (x_i, y_i, z_i)$ is the coordinate location of the $i$th ``puncture'',
solves~\eqref{eq:elec-constraint}.
Equation~(\ref{eq:ham-0}) then takes the form
\begin{equation}
\label{eq:ham-1}
  \triangle \psi + \frac{1}{4} \psi^{-3} \partial_i \varphi \ \partial_j \varphi \ \delta^{ij} = 0 \,.
\end{equation}

Following~\cite{Alcubierre:2009ij}, we now assume the following \emph{ansatz} for $\psi$
\begin{equation}
\label{eq:psi-ansatz}
    \psi^2 = \left( u + \eta \right)^2  - \frac{\varphi^2}{4} \,,
\end{equation}
where
\begin{equation}
\label{eq:eta-def}
\eta = \sum_{i=1}^{N}
        \frac{ m_i }{2|\mathbf{x} - \mathbf{x}_i|} \,.
\end{equation}

Equation~\eqref{eq:ham-1}, in terms of the new variable $u$, then reads
\begin{align}
  \triangle u & - \frac{\varphi^2}{4\psi^{2}(u+ \eta)}
                 \left(  \partial_k u \partial^k u 
                      + 2 \partial_k u \partial^k \eta 
                      + \partial_k \eta \partial^k \eta
                 \right) \notag \\
              & + \frac{\varphi}{2\psi^2} 
                 \left( \partial_k u \partial^k \varphi
                      + \partial_k \eta \partial^k \varphi
                 \right) \notag \\
              & + \frac{ 1 - (u + \eta)^2 }{4(u + \eta)\psi^2}
                 \partial_k \varphi \partial^k \varphi = 0 \,,
\label{eq:u-0}
\end{align}
where $\partial^k \equiv \delta^{kl} \partial_{l}$.
Note that when choosing configurations of BHs with the same charge-to-mass ratio, Eq.~(\ref{eq:u-0})
is immediately solved with $u=1$, and we recover the cases studied in Paper~I.

For our present purposes, we fix $m_1 = m_2 \equiv M/2$, $q_1 = -q_2 \equiv Q/2$ and $z_1 = -z_2 \equiv d/2$, and solve~\eqref{eq:u-0} by adapting the spectral solver \mbox{\textsc{TwoPunctures}}~\cite{Ansorg:2004ds}.
Originally developed to calculate four-dimensional vacuum puncture data corresponding to both single and binary BH configurations, 
\mbox{\textsc{TwoPunctures}} has been successfully adapted in the past to tackle different configurations (such as higher-dimensional puncture data~\cite{Zilhao:2011yc}). 
We here take a pragmatic approach to the initial data solving, and we merely modify the relevant source terms of the \mbox{\textsc{TwoPunctures}} routines according to~\eqref{eq:u-0}.

As noted in~\cite{Alcubierre:2009ij}, the function $u$ to solve for turns out to be only $C^0$ at the punctures, so the exponential convergence properties of the \mbox{\textsc{TwoPunctures}} solver are lost. 
This, however, is no concern
because in practice the numerical constraint violations are dominated
by the discretization errors accumulated in the time evolution such that
this relatively minor decrease in the accuracy of the initial data is
not noticeable in the evolution.
Furthermore we have cross-checked our initial data for several examples
with those obtained by the authors of~\cite{Alcubierre:2009ij} and found
very good agreement.\footnote{We thank J.C.~Degollado for these comparisons.}

The main physical observables we shall be interested in are the electromagnetic and gravitational radiation emitted in the collision process. To extract the radiation components we again follow the procedure described in Paper~I.  For the gravitational wave signal we calculate the Newman-Penrose
scalar $\Psi_4$ defined as
\begin{equation}
  \label{eq:Psi4}
  \Psi_4 \equiv C_{\alpha\beta\gamma\delta}
      k^{\alpha} \bar m^{\beta} k^\gamma \bar m^{\delta} \ ,
\end{equation}
where $C_{\alpha\beta\gamma\delta}$ is the Weyl tensor and $k$, $\bar
m$ are part of a null tetrad $l,k,m,\bar m$ satisfying $-l \cdot
k = 1 = m \cdot \bar m$; all other inner products vanish. 
For analyzing the behaviour of the electromagnetic fields we compute the scalar
functions $\Phi_1$ and $\Phi_2$~\cite{Newman:1961qr}, defined as
\begin{align}
  \label{eq:Phi1}
  \Phi_1 & \equiv \frac{1}{2} F_{\mu \nu} \left(
    l^{\mu} k^{\nu} + \bar m^{\mu} m^{\nu}
  \right) \ , \\
  \label{eq:Phi2}
  \Phi_{2} & \equiv F_{\mu \nu} \bar m^{\mu} k^{\nu} \ .
\end{align}
For outgoing waves at infinity, 
the relevant scalar behaves as
\begin{equation}
  \label{eq:Phi_asympt}
  \Phi_2 \sim E_{\hat \theta} - i E_{\hat \phi} \ .
\end{equation}
For static charges, the scalars behave as
\begin{equation}
  \label{eq:Phi_asympt_static}
  \Phi_1 \sim \frac{1}{2} E_{\hat r}, \qquad 
  \Phi_2 \sim  \frac{1}{2} \left( E_{\hat \theta} - i E_{\hat \phi} \right) \ .
\end{equation}

At a given extraction radius $R_\mathrm{ex}$, we perform a multipolar
decomposition by projecting $\Psi_4$, $\Phi_1$ and $\Phi_2$ onto spherical
harmonics of spin weight $s=-2$, $0$, and $-1$, respectively,
\begin{align}
  \Psi_4(t, \theta, \phi) & =
      \sum_{l,m} \psi^{lm}(t) Y_{lm}^{-2}(\theta,\phi) \ , \\
  \Phi_1(t, \theta, \phi) & =
      \sum_{l,m} \phi_{1}^{lm}(t) Y_{lm}^{0}(\theta,\phi) \ ,
      \label{eq:multipole_Phi1} \\
  \Phi_2(t, \theta, \phi) & =
      \sum_{l,m} \phi_{2}^{lm}(t) Y_{lm}^{-1}(\theta,\phi) \ .\label{eq:multipole_Phi2}
\end{align}
In terms of these multipoles, the radiated flux and energy are given by the
expressions~\cite{Newman:1961qr} 
\begin{align}
 \label{eq:GW-flux}
 P_{\rm GW} & = \frac{d E_{\rm GW}}{dt} =
     \lim_{r\to\infty} \frac{r^2}{16 \pi} \sum_{l,m}
     \left| \int_{-\infty}^t dt' \psi^{lm} (t') \right|^2 \ , \\
 P_{\rm EM} & = \frac{d E_{\rm EM}}{dt} =
     \lim_{r\to\infty} \frac{r^2}{4 \pi} \sum_{l,m} 
     \left|  \phi^{lm}_{2} (t) \right|^2 \ . \label{eq:EM-flux}
\end{align}

\section{Analytic predictions}
\label{classical_expectations}

Before presenting the results of our numerical simulations, we will discuss a simple analytic approximation to gain an intuitive understanding of the binary's dynamics. This analysis also provides predictions to compare with the numerical results presented below.

Consider the electrodynamics of a system of two oppositely charged point charges
in a Minkowski background spacetime. As in the BH case, we denote by
$q_1=-q_2\equiv Q/2$ and $m_1=m_2\equiv M/2$ the electric charge and mass of the
particles that are initially at rest at position $z=\pm d/2$.
The expected behaviour of the radial component of the resulting electric field
is given by~\cite{Jackson1998Classical}
\begin{equation}
  \label{eq:Er_asympt}
  E_{\hat r} = 4\pi\sum_{l=0}^{\infty}\sum_{m=-l}^l \frac{l+1}{2l+1} q_{lm}
      \frac{Y_{lm}(\theta,\varphi)}{r^{l+2}} \ ,
\end{equation}
where $q_{lm}=\int
Y_{lm}(\theta',\phi')^*(r')^l\rho(\mathbf{x}')d^3\mathbf{x}'$ are the
multipole moments and $\rho$ is the charge density.  The leading term
of this multipolar expansion for our system of two opposite charges is
the dipole
\begin{equation}
\label{eq:Er_2q}
E_{\hat r} \simeq \sqrt{\frac{4\pi}{3}} Qd \frac{Y_{10}}{r^3} \ ,
\end{equation}
whereas the monopole term vanishes because the total charge is
zero.
%
In a similar fashion, the $\theta$ component of the electric field becomes
\begin{equation}
\label{eq:Eth_2q}
E_{\hat \theta} \simeq - \sqrt{\frac{2\pi}{3}} Qd \frac{Y^{-1}_{10}}{r^3} \ .
\end{equation}

An estimate for the dipole amplitude in the limit
of two static point charges is then obtained from inserting the
radial and poloidal components of the electric field (\ref{eq:Er_2q}) and (\ref{eq:Eth_2q}) into
the expressions (\ref{eq:Phi_asympt_static}) for $\Phi_1$ and $\Phi_2$ and its
multipolar decompositions (\ref{eq:multipole_Phi1}) and (\ref{eq:multipole_Phi2})
\begin{align}
r^3\phi_1^{10} & =\sqrt{\frac{\pi}{3}} Q d\approx 1.02333 Qd\,,\label{eq:dipole}  \\
r^3\phi_2^{10} & = -\sqrt{\frac{\pi}{6}} Q d\approx - 0.72360 Qd \, .\label{eq:dipole-phi2} 
\end{align}
Although the actual setup is dynamical, one expects this expression to provide a reasonably good approximation
in the initial stages of the numerical evolution.
The comparison between this approximation and the numerical simulations will be performed below in Fig.~\ref{fig:phi-predict}.
After the merger and ringdown the dipole will eventually approach zero
as a single merged BH corresponds to the case $d=0$ in
Eqs.~(\ref{eq:dipole}) and (\ref{eq:dipole-phi2}).

\begin{figure*}[tpbh]
\centering
\includegraphics[width=0.45\textwidth]{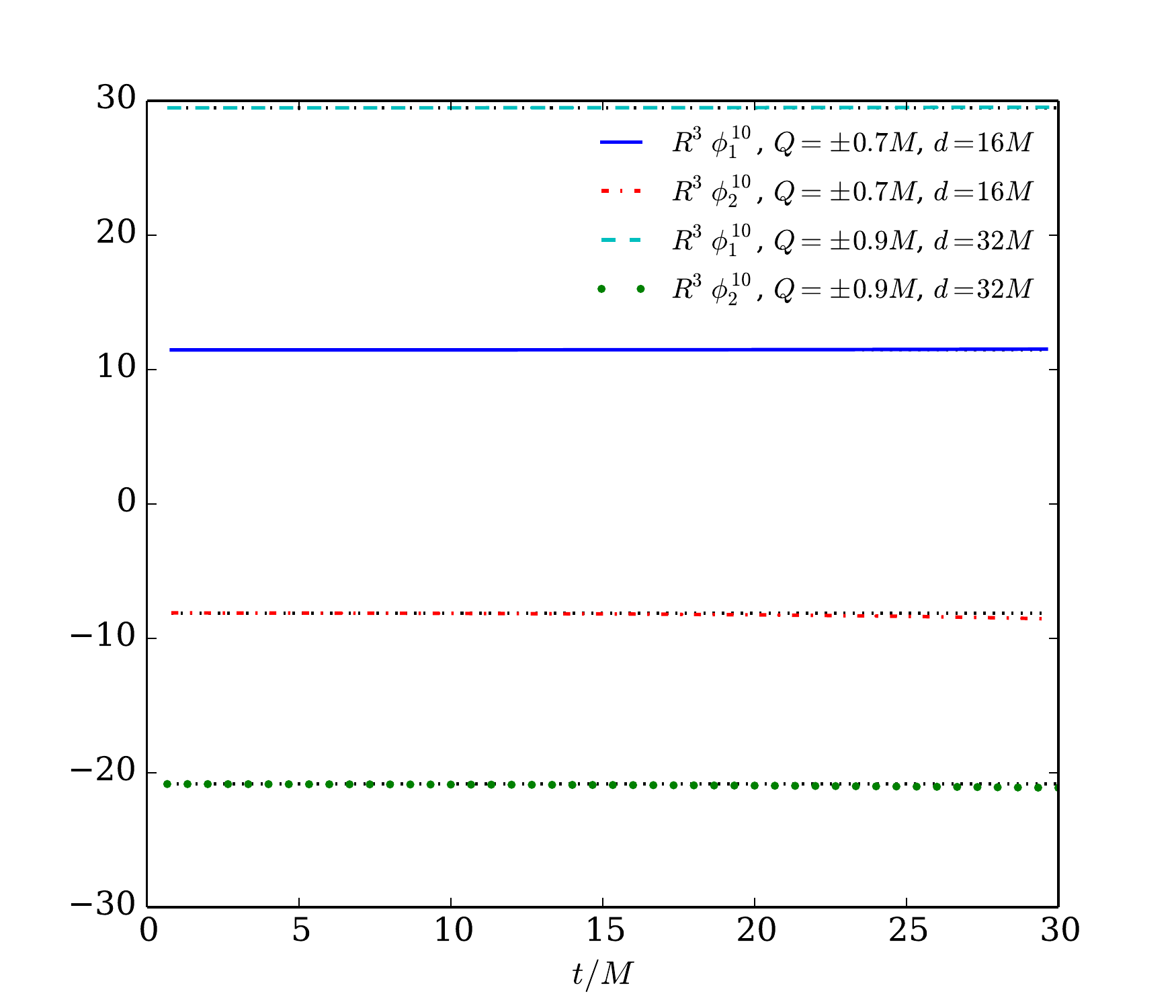}
\includegraphics[width=0.45\textwidth]{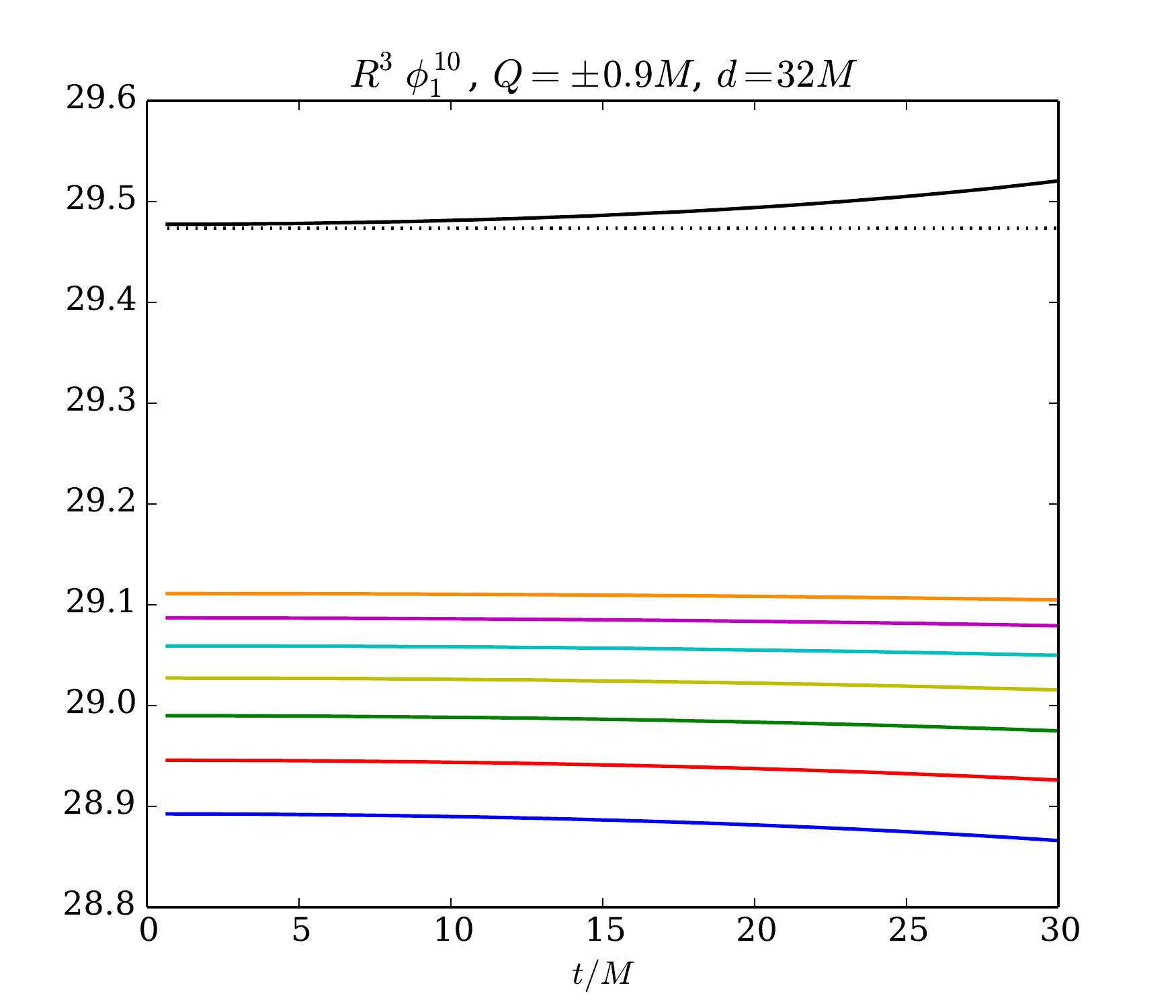}
\cprotect\caption[]{Predictions from Eqs.~\eqref{eq:dipole} and
  \eqref{eq:dipole-phi2} (black dotted lines) matched against our simulation
  results (models \verb|q+-070_d16_hf80| and \verb|q+-090_d32_hf192|; see
  Table~\ref{tab:runs}). Left plot shows results obtained from fitting a curve
  of the form $R_{\rm ex}^{3} \phi_{1,2}^{10} = a_0 + a_1/R_{\rm ex}$ to the
  numerically extracted $\phi_{1,2}^{10}$ (for all time steps) using all
  available extraction radii.  Right plot (showing only the $Q=\pm 0.9M$ case)
  depicts explicitly the extracted $R_{\rm ex}^{3} \phi_{1}^{10}$ for all
  extraction radii. In this case, the curves from bottom to top correspond to
  $R_{\rm ex}=100M,\ldots, 160M$ in steps of $10M$, with the uppermost being the
  result extrapolated to infinity.  We note that for these cases, the BH
  merger happens roughly at around $t\sim 330M$. \label{fig:phi-predict}}
\end{figure*}

Let us now follow the non-relativistic dynamics of the two charges in Minkowski
spacetime assuming their interaction is encoded in the Newtonian gravitational
energy plus the electrostatic energy:
\begin{equation}
 V=-\frac{GM^2}{4d}-\frac{1}{16\pi\epsilon_0}\frac{Q^2}{d}\ .
\end{equation}
%
Our aim is to obtain the radiated energy in the collision using dipole/quadrupole emission formulas for electromagnetic/gravitational radiation. This is expected to be a good approximation for systems where the accelerations involved are not too large and it has yielded good estimates in the case of equal charges, as shown in Paper~I. Using units with $G=4\pi \epsilon_0 = 1$, conservation of energy implies that under their mutual attraction the motion along the $z$ axis obeys
\begin{equation}
  M\dot{z}^2-\frac{M^2{\cal B}}{4z}=-\frac{M^2{\cal B}}{2d} \ , \label{eqm2}
\end{equation}
%
where
\begin{equation}
{\cal B}\equiv 1+Q^2/M^2\ .
\end{equation}
The resulting equation of motion for $z(t)$ is obtained by differentiating
Eq.~(\ref{eqm2}) which results in
\begin{equation}
  M\ddot{z}=-\frac{M^2}{8z^2}-\frac{Q^2}{8z^2}=
      -M^2\frac{{\cal B}}{8z^2}\ . \label{eqm1}
\end{equation}


To compare the emission of gravitational and electromagnetic radiation we use the quadrupole formula for the total power emitted in gravitational radiation
\begin{equation}
P_{\mathrm{GW}}=\frac{G}{45c^5}\sum_{ij}(\dddot{Q}_{ij})^2 \ ,
\label{quadrupole}
\end{equation}
where the (traceless) quadrupole tensor $Q_{ij}=\int d^3\mathbf{x} \ \rho_m(\mathbf{x})(3x_ix_j-r^2\delta_{ij})$, where $\rho_m$ is the matter energy density,
and the Larmor dipole formula for the total power emitted in electromagnetic radiation by a time-varying dipole~\cite{Jackson1998Classical}
\begin{equation}
P_{\rm EM}=\frac{1}{6\pi\epsilon_0 c^3}\ddot{\bf d}^2\ ,\label{eq:power}
\end{equation}
where ${\bf d}$ is the dipole vector that has components $ d_i=\int d^{3}\mathbf{x} \ \rho_e(\mathbf{x})x_i$, and $\rho_e$ is the electric charge density. For clarity we have reinstated the factors $G,c$ and $4\pi \epsilon_0$ in the last two formulas. We shall drop them in the following. 
To compute \eqref{quadrupole} and  \eqref{eq:power} we  use \eqref{eqm2} and  \eqref{eqm1} to find
\begin{equation}
P_{\rm GW}=\frac{{\cal B}^3M^5}{480z^4}\left(\frac{1}{z}-\frac{2}{d}\right) \ , 
\qquad P_{\rm EM}=\frac{{\cal B}^2M^2Q^2}{96z^4}\ .
\end{equation}

Using $\int dt (\cdots ) = \int dz/\dot{z} (\cdots)$, we can evaluate
the time integral up to some cutoff separation, say $z_c$. 
This gives
\begin{widetext}
\begin{equation}
\label{GWprediction}
\frac{E^{\rm GW}_{\rm rad}}{M} ={\cal B}^{5/2}M^{7/2} \frac{15d^2+24dz_c+32z_c^2}{12600(dz_c)^{2}}\left(\frac{1}{z_c}-\frac{2}{d}\right)^{3/2}\ \ \ \stackrel{d\rightarrow \infty}{\longrightarrow} \ \ \  \frac{\displaystyle{\left(1+\frac{Q^2}{M^2}\right)^{5/2}}}{\displaystyle{840\left(\frac{z_c}{M}\right)^{7/2}}} ,
\end{equation}
and
\begin{equation}
\label{EMdipprediction}
\frac{E^{\rm EM}_{\rm rad}}{M} ={\cal B}^{3/2}M^{1/2}Q^2 \frac{3d^2+8dz_c+32z_c^2}{360(dz_c)^{2}}\sqrt{\frac{1}{z_c}-\frac{2}{d}}\ \ \ \stackrel{d\rightarrow \infty}{\longrightarrow} \ \ \  \frac{\displaystyle{\left(1+\frac{Q^2}{M^2}\right)^{3/2}\left(\frac{Q}{M}\right)^2}}{\displaystyle{120\left(\frac{z_c}{M}\right)^{5/2}}} .
\end{equation}
\end{widetext}
Thus, for large initial separations,
\begin{equation}
\frac{E^{\rm EM}_{\rm rad}}{E^{\rm GW}_{\rm rad}}=7\frac{z_c}{M} \frac{\displaystyle{\left(\frac{Q}{M}\right)^2}}{\displaystyle{1+\frac{Q^2}{M^2}}}\ .
\label{prediction_ratio}
\end{equation}

We can now make some estimates based on the previous formulas in the limit of $d\rightarrow \infty$. These estimates depend on the cutoff scale $z_c$. 
As we will see in the next section, we observe that indeed, for a range of cutoffs around $z_c \simeq 1.5M$, the analytic approximation captures remarkably well the radiation emission patterns, for both the electromagnetic and the gravitational wave sectors.
At the end of the next section we shall make some explicit comparisons between the above formulas and the numerical results.

\section{Numerical results}
\label{sec:numerical_results}

\begin{table*}[ht]
  \centering
  \caption{Numerical grid structure used (in the notation of Sec.~II~E
    of~\cite{Sperhake:2006cy}), initial coordinate distance $d/M$, total Arnowitt, Deser, Misner (ADM) mass,
    charge-to-mass ratio $\pm Q/M$, gravitational ($E_{\rm rad}^{\mathrm{GW}}$)
    and electromagnetic ($E_{\rm rad}^{\mathrm{EM}}$) radiated energy for our set
    of simulations. Gravitational radiated energy has been computed using only the
    $l=2$, $m=0$ mode, while for the electromagnetic radiated energy only the
    $l=1$, $m=0$ multipole was used, as the energy contained in higher-order
    multipoles is negligible for all configurations. \label{tab:runs}}
  
\begin{tabular*}{\textwidth}{@{\extracolsep{\fill}}lcccccc}
\hline 
\hline
Run  & 	  Grid structure  & 	  $M_{\mathrm{ADM}}$ & 	  $d/M$  & 	  $ |Q|/M$ &  	 $\frac{E_{\mathrm{rad}}^{\mathrm{GW}}}{M_{\mathrm{ADM}}}  \times 10^{3}$  & 	 $\frac{E_{\mathrm{rad}}^{\mathrm{EM}}}{M_{\mathrm{ADM}}}  \times 10^{3}$ \\
\hline
\verb|q+-010_d16_hf64|  & 	 $\{(256,128,64,32,16\times(4,2,1,0.5), M/64\}$  & 	 1  & 	 16  & 	 0.1  & 	 0.536  & 	 0.0426 \\ 
\verb|q+-020_d16_hf64|  & 	 $\{(256,128,64,32,16\times(4,2,1,0.5), M/64\}$  & 	 0.999  & 	 16  & 	 0.2  & 	 0.554  & 	 0.174 \\ 
\verb|q+-030_d16_hf64|  & 	 $\{(256,128,64,32,16\times(4,2,1,0.5), M/64\}$  & 	 0.997  & 	 16  & 	 0.3  & 	 0.584  & 	 0.405 \\ 
\verb|q+-040_d16_hf64|  & 	 $\{(256,128,64,32,16\times(4,2,1,0.5), M/64\}$  & 	 0.995  & 	 16  & 	 0.4  & 	 0.627  & 	 0.754 \\ 
\verb|q+-050_d16_hf64|  & 	 $\{(256,128,64,32,16\times(4,2,1,0.5), M/64\}$  & 	 0.993  & 	 16  & 	 0.5  & 	 0.685  & 	 1.25 \\ 
\verb|q+-050_d16_hf80|  & 	 $\{(256,128,64,32,16\times(4,2,1,0.5), M/80\}$  & 	 0.993  & 	 16  & 	 0.5  & 	 0.706  & 	 1.26 \\ 
\verb|q+-050_d16_hf96|  & 	 $\{(256,128,64,32,16\times(4,2,1,0.5), M/96\}$  & 	 0.993  & 	 16  & 	 0.5  & 	 0.714  & 	 1.26 \\ 
\verb|q+-060_d16_hf64|  & 	 $\{(256,128,64,32,16\times(4,2,1,0.5), M/64\}$  & 	 0.989  & 	 16  & 	 0.6  & 	 0.757  & 	 1.92 \\ 
\verb|q+-070_d16_hf64|  & 	 $\{(256,128,64,32,16\times(4,2,1,0.5), M/64\}$  & 	 0.985  & 	 16  & 	 0.7  & 	 0.846  & 	 2.82 \\ 
\verb|q+-070_d16_hf80|  & 	 $\{(256,128,64,32,16\times(4,2,1,0.5), M/80\}$  & 	 0.985  & 	 16  & 	 0.7  & 	 0.875  & 	 2.84 \\ 
\verb|q+-070_d16_hf96|  & 	 $\{(256,128,64,32,16\times(4,2,1,0.5), M/96\}$  & 	 0.985  & 	 16  & 	 0.7  & 	 0.885  & 	 2.84 \\ 
\verb|q+-080_d16_hf64|  & 	 $\{(256,128,64,32,16\times(4,2,1,0.5), M/64\}$  & 	 0.981  & 	 16  & 	 0.8  & 	 0.953  & 	 4 \\ 
\verb|q+-090_d16_hf64|  & 	 $\{(256,128,64,32,16\times(4,2,1,0.5), M/64\}$  & 	 0.976  & 	 16  & 	 0.9  & 	 1.08  & 	 5.52 \\ 
\verb|q+-090_d16_hf80|  & 	 $\{(256,128,64,32,16\times(4,2,1,0.5), M/80\}$  & 	 0.976  & 	 16  & 	 0.9  & 	 1.12  & 	 5.58 \\ 
\verb|q+-090_d16_hf96|  & 	 $\{(256,128,64,32,16\times(4,2,1,0.5), M/96\}$  & 	 0.976  & 	 16  & 	 0.9  & 	 1.13  & 	 5.59 \\ 
\verb|q+-050_d32_hf96|  & 	 $\{(256,176,64,32\times(8,4,2,1,0.5), M/96\}$  & 	 0.996  & 	 32  & 	 0.5  & 	 0.755  & 	 1.35 \\ 
\verb|q+-060_d32_hf96|  & 	 $\{(256,176,64,32\times(8,4,2,1,0.5), M/96\}$  & 	 0.995  & 	 32  & 	 0.6  & 	 0.84  & 	 2.08 \\ 
\verb|q+-070_d32_hf96|  & 	 $\{(256,176,64,32\times(8,4,2,1,0.5), M/96\}$  & 	 0.993  & 	 32  & 	 0.7  & 	 0.945  & 	 3.05 \\ 
\verb|q+-080_d32_hf96|  & 	 $\{(256,176,64,32\times(8,4,2,1,0.5), M/96\}$  & 	 0.99  & 	 32  & 	 0.8  & 	 1.07  & 	 4.32 \\ 
\verb|q+-090_d32_hf128|  & 	 $\{(256,176,64,32\times(8,4,2,1,0.5,0.25), M/128\}$  & 	 0.988  & 	 32  & 	 0.9  & 	 1.16  & 	 5.92 \\ 
\verb|q+-090_d32_hf160|  & 	 $\{(256,176,64,32\times(8,4,2,1,0.5,0.25), M/160\}$  & 	 0.988  & 	 32  & 	 0.9  & 	 1.2  & 	 5.95 \\ 
\verb|q+-090_d32_hf192|  & 	 $\{(256,176,64,32\times(8,4,2,1,0.5,0.25), M/192\}$  & 	 0.988  & 	 32  & 	 0.9  & 	 1.22  & 	 5.97 \\ 
\verb|q+-093_d32_hf192|  & 	 $\{(256,176,64,32\times(8,4,2,1,0.5,0.25), M/192\}$  & 	 0.987  & 	 32  & 	 0.93  & 	 1.27  & 	 6.54 \\ 
\verb|q+-095_d32_hf256|  & 	 $\{(256,176,64,32\times(8,4,2,1,0.5,0.25,0.125), M/256\}$  & 	 0.986  & 	 32  & 	 0.95  & 	 1.24  & 	 6.89 \\ 
\verb|q+-095_d32_hf320|  & 	 $\{(256,176,64,32\times(8,4,2,1,0.5,0.25,0.125), M/320\}$  & 	 0.986  & 	 32  & 	 0.95  & 	 1.29  & 	 6.94 \\ 
\verb|q+-097_d32_hf320|  & 	 $\{(256,176,64,32\times(8,4,2,1,0.5,0.25,0.125), M/320\}$  & 	 0.986  & 	 32  & 	 0.97  & 	 1.32  & 	 7.36 \\ 
\verb|q+-099_d32_hf512|  & 	 $\{(256,176,64,32\times(8,4,2,1,0.5,0.25,0.125,0.0625), M/512\}$  & 	 0.985  & 	 32  & 	 0.99  & 	 1.32  & 	 7.79 \\ 
\verb|q+-099_d32_hf640|  & 	 $\{(256,176,64,32\times(8,4,2,1,0.5,0.25,0.125,0.0625), M/640\}$  & 	 0.985  & 	 32  & 	 0.99  & 	 1.36  & 	 7.82 \\ 
\verb|q+-090_d48_hf160|  & 	 $\{(256,176,64\times(16,8,4,2,1,0.5,0.25), M/160\}$  & 	 0.992  & 	 48  & 	 0.9  & 	 1.2  & 	 6.01 \\ 
\hline
\hline
\end{tabular*}

\end{table*}

As in Paper~I, we numerically integrate the
Einstein-Maxwell system
using fourth-order spatial discretization with the \textsc{Lean} code, which is
based on the \textsc{Cactus} Computational toolkit~\cite{cactus}, the
\textsc{Carpet} mesh refinement package~\cite{Schnetter:2003rb,carpet} and uses
\textsc{AHFinderDirect} for tracking apparent
horizons~\cite{Thornburg:2003sf,Thornburg:1995cp}. \textsc{Lean} uses the Baumgarte, Shapiro, Shibata, Nakamura formulation of the Einstein equations~\cite{Shibata:1995we,Baumgarte:1998te}
with the moving puncture method~\cite{Campanelli:2005dd,Baker:2005vv}. We refer
the interested reader to Ref.~\cite{Sperhake:2006cy} for further details on the
numerical methods, and to Paper~I for the tests performed with
the Einstein-Maxwell implementation.

\subsection{Simulations and convergence properties}

As stated above, we have prepared time-symmetric binary BH puncture-type
initial data with $m_1=m_2=M/2$, $q_1=-q_2=Q/2$, where we vary the
charge-to-mass
ratio from $Q/M= \pm 0.1$ to $Q/M= \pm 0.99 $. Binaries start from rest with
initial (coordinate) distance $d/M = 16$, $d/M = 32$ or $d/M = 48$.
These parameters as well as the grid setup and the radiated
energy emitted in gravitational ($E_{\rm rad}^{\rm GW}$)
and electromagnetic ($E_{\rm rad}^{\rm EM}$) waves are listed in
Table~\ref{tab:runs}.
In describing the
grid structure, we follow the notation of Sec.~II~E of~\cite{Sperhake:2006cy}:
the initial grid consists of $n$ outer levels
centered on the origin (remaining static throughout the simulation) and $m$
moving levels with two components centered around each BH; for example,
$\{(256,128,64,32,16\times(4,2,1,0.5), M/64\}$ specifies a grid with five fixed
outer components of radii 256, 128, 64, 32 and 16, and four moving levels with
radii 4, 2, 1 and 0.5.  The grid spacing is $h_f = M/64$ on the finer level and
successively increases by factors of 2 until the outermost level.

As a test on the correctness of our implementation, we have evolved models 
\verb|q+-050_d16| and \verb|q+-090_d32| with three different resolutions, as
outlined in Table~\ref{tab:runs}, and performed a convergence analysis
of both the extracted waveforms and the violation of the electric and
Einstein constraints.
\begin{figure*}[tbh]
\centering
\includegraphics[width=0.45\textwidth,clip=true]{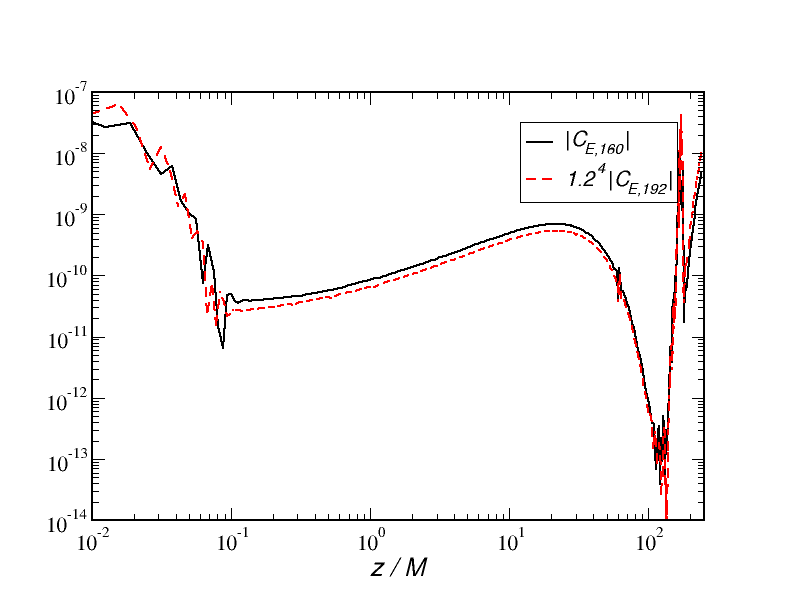}
\includegraphics[width=0.45\textwidth,clip=true]{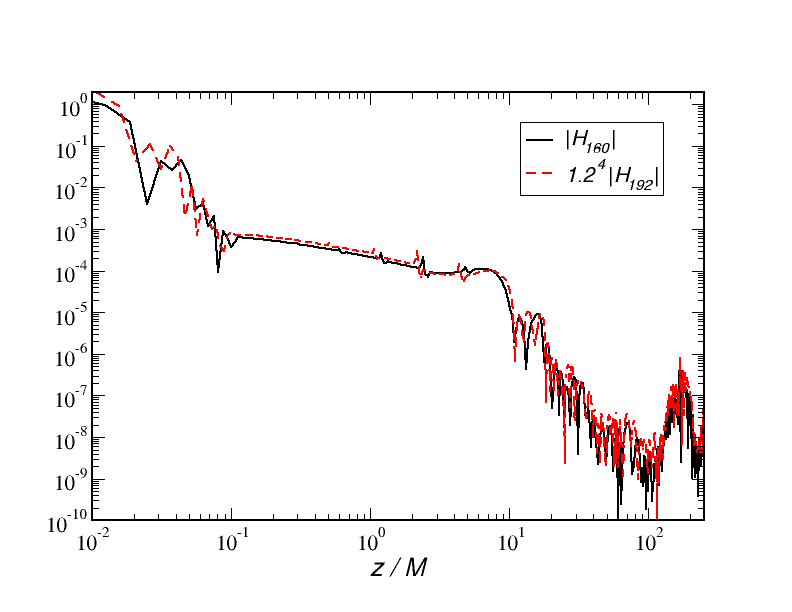}
\cprotect\caption{The electric (left panel) and Hamiltonian (right panel)
constraints along the collision axis at time $t=384M$ for model
\verb|q+-090_d32|. The solid (black)
curves display the result obtained for lower resolution $h_f=M/160$
and the dashed (red) curves show that obtained for higher resolution
$h_f=M/196$ and amplified by $1.2^4$ for the expected fourth-order
convergence. \label{fig:conv_EH}}
\end{figure*}

In Fig.~\ref{fig:conv_EH}, we display the electric and Hamiltonian
constraints along the $z$ axis at $t=384M$ for model \verb|q+-090_d32|. Each panel displays two curves,
one obtained for a resolution $h_f=M/160$ on the finest grid and one
obtained for a higher resolution $h_f=M/192$ for which the result has been
amplified by the expected convergence factor $1.2^4$. Even though some
constraint violations are generated by the outer boundary treatment,
the overall violations are small (as compared with the magnitude of the
individual terms summed over in the constraints)
and display fourth-order convergence.
For the momentum constraints we observe similar behaviour to that
of the Hamiltonian constraint.

We have complemented these two simulations with an additional one using
$h_f=M/128$ in order to estimate the uncertainties in the energy
radiated in gravitational and electromagnetic waves. We obtain for
the electromagnetic energy $E_{\rm rad}^{\rm EM}$ the values
$0.5865\%~M$, $0.5883\%~M$ and $0.5888\%~M$
respectively for
$h_f=M/128$, $1/160$ and $1/192$ which is in good agreement with
fourth-order convergence and gives a relative error of half a percent
or better for $h_f \le M/128$. Likewise, we obtain for the energy
emitted in gravitational waves $E_{\rm rad}^{\rm GW} = 0.1189\%~M$,
$0.1214\%~M$ and $0.1222\%~M$, again in good agreement with
fourth-order convergence, and yielding a relative error of about
$3\%$ for $h_f=M/128$ and $1.5\%$ or less for $h_f\le M/160$.
We obtain similar error estimates for the model \verb|q+-050_d16|
using the three resolutions $h_f=M/64$, $h_f=M/80$ and $h_f=M/96$.
This observation confirms our expectation that accurate evolutions
with larger magnitudes of the electric charge require higher numerical
resolution. We also monitored the uncertainties in the radiated
energies arising from extraction at finite radius by extrapolating the
results to infinity. For this purpose, we have extracted the
wave signals at $R_{\rm ex}/M=80,\,100,\,110,\,120,\,130,\,140,\,150,$
and $160$. By extrapolating the values obtained at these
finite radii to infinity using a $a_0 + a_1/R_{\rm ex}$ dependence,
we determine
the relative uncertainties at $R_{\rm ex} = 160~M$ to be about $0.5\%$ for
the electromagnetic and $1\%$ for the gravitational wave energy
radiated away from the binary. Unless stated otherwise, the reported
energies refer to the extrapolated values. 
A further uncertainty in our results
arises from the finite initial separation of the BHs. We estimate the
resulting error by studying collisions for $Q/M=\pm 0.9$ starting from
separations $d/M=16$, $32$ and $48$. As intuitively expected, the radiated
energies mildly increase with initial separation.
By extrapolating the results
to infinite $d$ assuming a $a_0 + a_1/d$ dependence, we obtain a numerical
uncertainty of about $5\%$ for our values obtained for $d/M=32$. Combining
all three sources of errors, we estimate the total uncertainty to be
$6\%$ for $E_{\rm rad}^{\rm EM}$ and $7.5\%$ for $E_{\rm rad}^{\rm GW}$. If interpreted
as energy radiated by head-on collisions starting at finite separation,
these uncertainty estimates drop by a factor of about 3.

\subsection{Waveforms and integrated energy}
%
\begin{figure*}[tbph]
\centering
\includegraphics[width=0.45\textwidth]{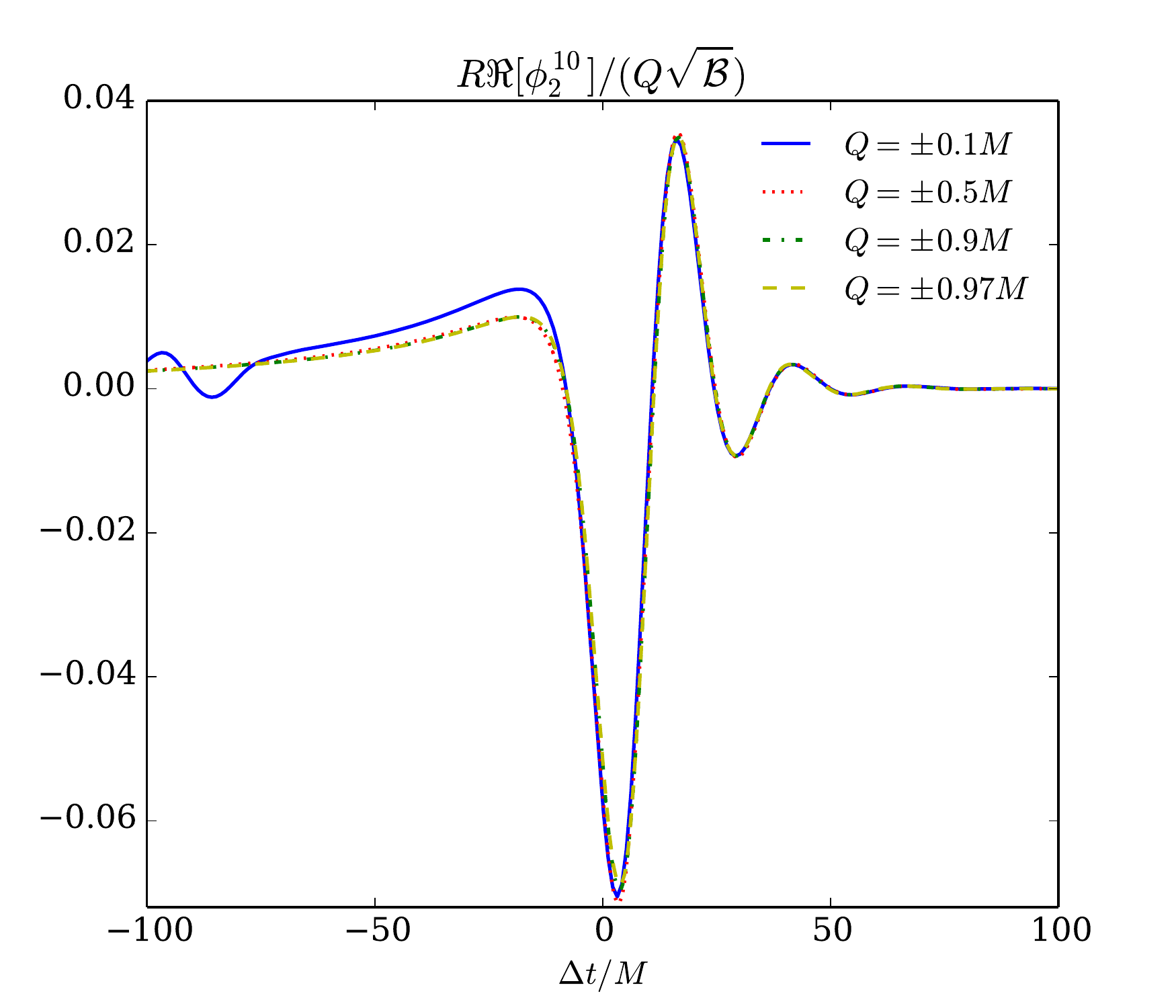}
\includegraphics[width=0.45\textwidth]{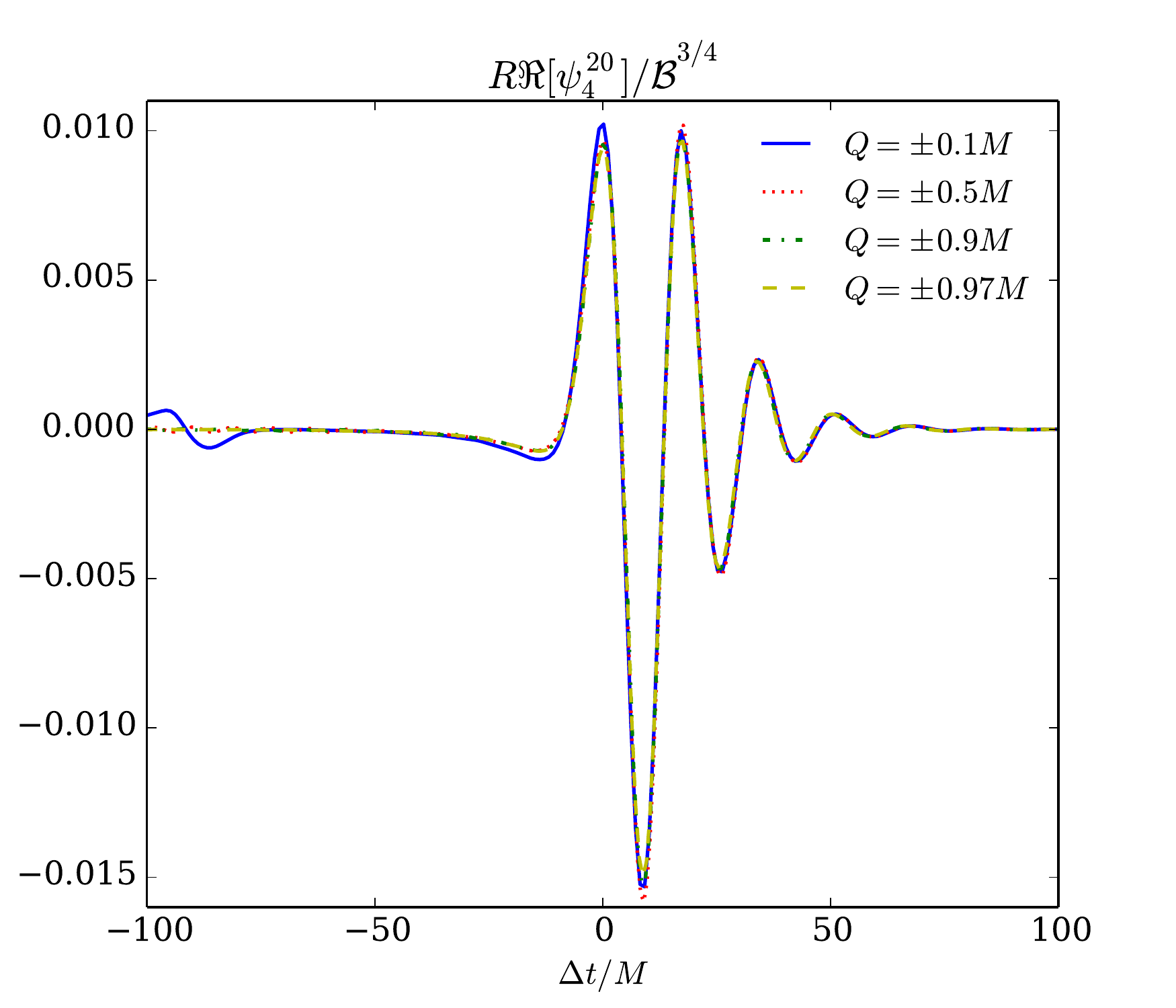}
\caption{Real part of the electromagnetic ($l=1$, $m=0$ mode) and gravitational
  ($l=2$, $m=0$ mode) waveforms. These have been conveniently rescaled and shifted in
  time so that their peaks coincide. \label{fig:waveforms}}
\end{figure*}
In Fig.~\ref{fig:waveforms} we display the waveforms $\psi^{20}_4$ and
$\phi^{10}_2$ obtained at finite extraction radius for a representative subset
of the initial configurations listed in Table~\ref{tab:runs}.

Note that the configurations studied in this work differ qualitatively
from those of Paper~I in the final outcome of the
merger: a charged BH in Paper~I but an electrically
neutral, i.e.~a Schwarzschild BH, in this study.
In consequence, the ringdown of the collisions in Paper~I exhibits a superposition of both
gravitational and electromagnetic quasi-normal modes (QNM) in both the
$\psi^{20}_4$ and $\phi^{10}_2$ waveforms. For the electrically
neutral post-merger BHs of this study, in contrast,
the gravitational wave signal $\psi^{20}_4$ matches the
ringdown of the neutral Schwarzschild BH and we find a strong \emph{electromagnetic}
QNM component in $\phi^{10}_2$. We find no signs of mixing between electromagnetic and gravitational
modes. For instance, for all configurations of Table
\ref{tab:runs}, the $\phi^{10}_2$ waveform is very well described by the lowest
electromagnetic ringdown mode~\cite{Berti:2005ys,Berti:2009kk}. In fact
one can recover the lowest electromagnetic QNM of Schwarzschild BHs, as
given by perturbative studies, with an accuracy of $0.5\%$.

We have further found that the
dependency of the multipoles $\psi^{20}_4$ and $\phi^{10}_2$
on the parameters $\cal{B}$ and $Q$ is very well modeled by the following
simple scaling laws: $\psi^{20}_4 \sim {\cal B}^{3/4}$ and
$\phi^{10}_2 \sim Q \sqrt{\cal B}$.
For the oppositely charged binaries of the present study, this scaling
appears to be satisfied with an even higher accuracy than in the
equal-charge case displayed in Fig.~5 of Paper~I.
Possibly this is a consequence of the vanishing charge of
the final BH. We indeed observe that most of the waveform signal
is emitted after formation of the common apparent horizon and would
therefore be expected to carry the
signature of the final BH.
While the ringdown frequency is
determined by the quasi-normal ringing of a neutral BH, it is
interesting to note that the amplitude can be recovered using the above
scaling laws. In consequence, the knowledge of the
$\psi^{20}_4$ and $\phi^{10}_2$ multipoles for a single charge-to-mass ratio
allows us to derive by rescaling
the corresponding waveforms for \emph{any} other charge-to-mass ratio
without the need of performing any other numerical evolution.

\begin{figure}[htbp]
\centering
\includegraphics[width=0.47\textwidth]{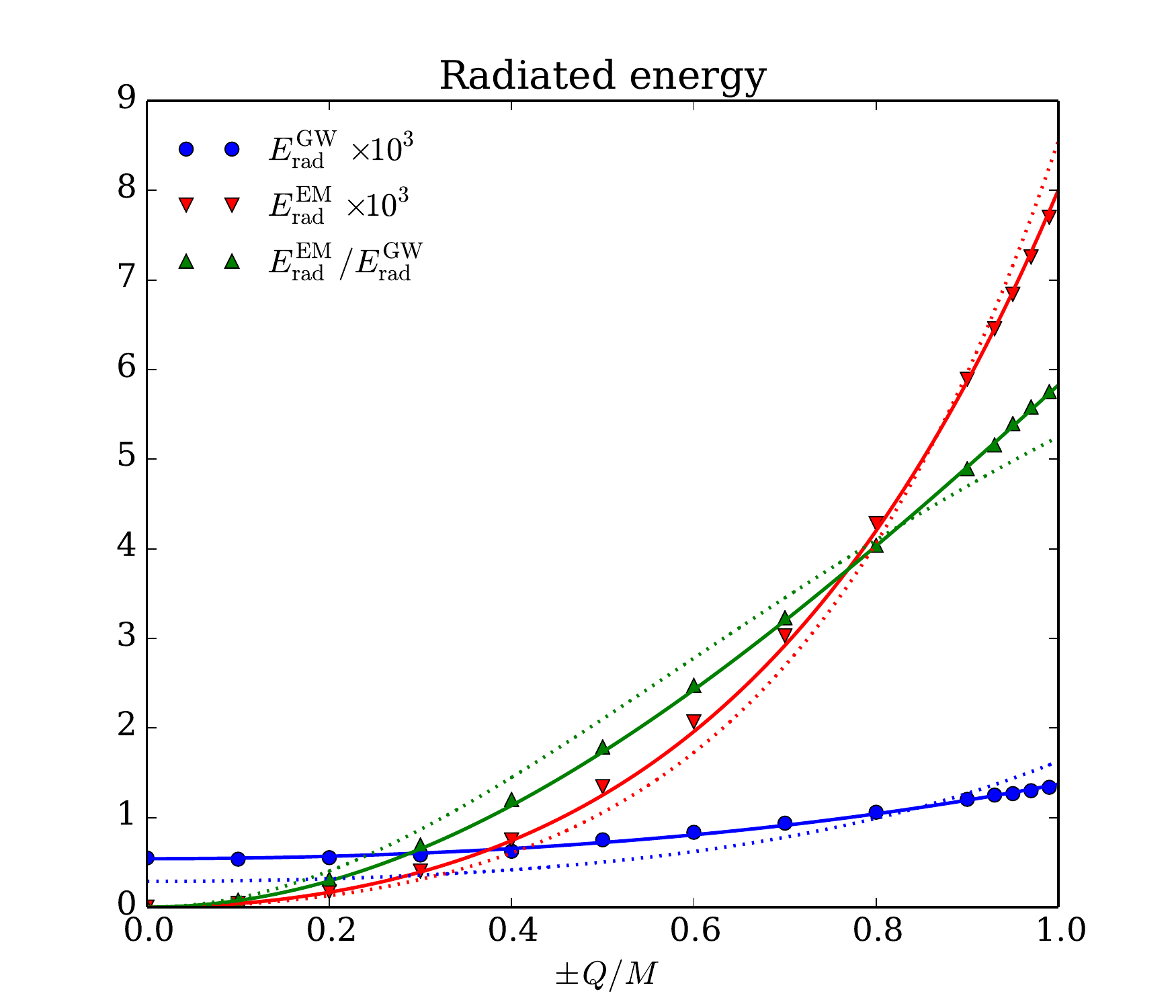}
\caption{Total energy radiated in the electromagnetic
  ($E^{\rm EM}_{\rm rad}$) and gravitational ($E^{\rm GW}_{\rm rad}$)
  channels. Solid lines show a fit to the numerical results of the form $E_{\rm
    rad}^{\rm GW} = 8.53 \times 10^{-5} + 4.55 \times 10^{-4} \ {\cal B}^{3/2}$,
  $E_{\rm rad}^{\rm EM} = 4.00 \times 10^{-3} \ Q^2 {\cal B}$, in agreement with
  the scaling used in Fig.~\ref{fig:waveforms}.  Dotted lines show results from
  the analytic approximation taking $z_c/M=1.5$. \label{fig:energy} }
\end{figure}

We compute the total radiated energies according to
Eqs.~(\ref{eq:GW-flux}) and (\ref{eq:EM-flux}). To account for
spurious, unphysical radiation resulting from the
initial data construction, we start
the integration of the radiated fluxes at some finite time $\Delta t$ after the
start of the simulation, thus allowing the spurious pulse to first radiate off
the computational domain. In practice, we find a value
$\Delta t = R_{\rm ex}+50~M$ to be sufficient for this purpose.
The radiated energies thus obtained are plotted in Fig.~\ref{fig:energy} as
functions of the charge-to-mass ratio and quantitatively illustrate the scaling
discussed in the previous paragraph.

These results contrast with the corresponding equal charge collisions of
Paper~I, where the emitted gravitational radiation decreases
with increasing charge because of its decelerating effect and the
correspondingly low collision velocities,
and the emitted electromagnetic radiation peaks at around $Q/M=0.6$. In the
case of opposite charges, in contrast,
both gravitational and electromagnetic radiation
increase with $Q/M$, and the electromagnetic radiation becomes
the dominating channel for $|Q|/M \gtrsim 0.37$.

As already mentioned at the end of the last section, we
observe a good agreement between our simple analytic model of
Sec.~\ref{classical_expectations} and the numerical simulations we
have just presented, for a range of cut-offs for the former. For
instance, setting $z_c=1.5 M$, we make the following observations:
(i) For the configuration $|Q|/M=0.99$, the ratio of
energy in electromagnetic to gravitational radiation obtained in our
numerical simulations is $\sim 5.8$ (cf.~Table \ref{tab:runs}),
whereas that obtained from our simple analytic approximation is
$5.2$; cf.~Eq.~(\ref{prediction_ratio}). (ii) Equal amounts of
electromagnetic and gravitational radiated energies are obtained
for $|Q|/M\sim 0.37$ in the numerical simulations and $|Q|/M= 0.31$ in the
analytic model. The analytical results thus reproduce the
numerical values with an error between $10\%$ and $20\%$.
A comparison of the energies emitted in gravitational and electromagnetic
radiation for the entire range $Q/M$
is shown in Fig.~\ref{fig:energy}.
Even though a discrepancy at a level of about $10\%$
is visible, the analytic prediction captures the main features of the
energy emission remarkably well.

Cutoff independent estimates
are provided by the ratio of the energy emitted in either gravitational or electromagnetic waves for two different values of the charge. For instance, from the numerical simulations
\begin{equation}
\frac{E^{\rm GW}_{\rm rad}(|Q|=0.99)}{E^{\rm GW}_{\rm rad}(Q=0)}\sim 2.7 \, \qquad \frac{E^{\rm EM}_{\rm rad}(|Q|=0.99)}{E^{\rm EM}_{\rm rad}(Q=0.1)}\sim 184 \ ,
\end{equation}
whereas the corresponding values from the analytical approximation are, from (\ref{GWprediction}) and (\ref{EMdipprediction}), 
\begin{equation}
\frac{E^{\rm GW}_{\rm rad}(|Q|=0.99)}{E^{\rm GW}_{\rm rad}(Q=0)}\sim5.6 \, \qquad \frac{E^{\rm EM}_{\rm rad}(|Q|=0.99)}{E^{\rm EM}_{\rm rad}(Q=0.1)}\sim 269 \ ,
\end{equation}
corresponding to mismatches of $\sim 2$ and $\sim 1.5$ respectively.

\section{Conclusions}
\label{sec:final}
The number of applications of numerical relativity to high energy physics has
been growing enormously in recent years \cite{Cardoso:2012qm}. One particular
line of research in this area has been the understanding of high energy
collisions of BHs
\cite{Sperhake:2008ga,Shibata:2008rq,Sperhake:2010uv,Sperhake:2012me} and other
compact objects~\cite{Choptuik:2009ww,East:2012mb} and a main open question in
this context concerns the impact of electric charge on the collision dynamics.
To address this question we have continued in this paper the programme
initiated in Paper~I of studying charged BH collisions.  We have here focused on
oppositely charged BHs with the same mass---as to maximize the acceleration of
the system and hence the gravitational and electromagnetic wave emission---and
have shown that the numerical simulations and the extraction of the observable
quantities---gravitational and electromagnetic radiation---are well under
control.

We have successfully evolved configurations with $|Q|/M$ ranging from 0.1 to
0.99, once again showing that cases with nearly extremal charge, albeit
requiring higher numerical resolution, are simpler to model numerically than the
corresponding nearly extremal spin cases~\cite{Lovelace:2011nu,Lousto:2012es}.

The observed qualitative behaviour of the energy radiated away is summarized in
Fig.~\ref{fig:energy} and demonstrates that both electromagnetic and
gravitational radiation increase monotonically as the (opposite) charges are
increased. More surprisingly, our study has revealed a simple, apparently
universal scaling of the energy dependence on the charge magnitude that can be
seen both in Fig.~\ref{fig:energy} and in the waveforms presented in
Fig.~\ref{fig:waveforms}. This scaling suggests that the head-on collision of
charged BHs with opposite charge to mass ratios may have a (hidden) conformal
symmetry, a possibility deserving further study. We have further shown that the
radiation emission is well described by a simple analytical model of two
non-relativistic charges in Minkowski space. The radiation emission predicted
for the gravitational quadrupole and electromagnetic dipole time variations by
numerical relativity calculations and by analytic methods show good agreement in
Fig.~\ref{fig:energy}.

There are two natural extensions of this study. One is to perform high energy collisions of
charged BHs.  Introducing non-zero boosts into the initial data, however,
represents a non-trivial challenge since the full (coupled) system of constraint
equations needs to be solved.  Work in this direction is underway.

Non head-on collisions or binaries in quasi-circular orbits are another natural extension of our results.
This problem is understood for neutral, spinning binaries, where it was observed that the total radiation output in the process
increases for larger {\it final} black hole spins. In simple terms, this is because the innermost stable
circular orbit (ISCO) moves inwards and the binary can sweep higher frequencies and radiate more strongly. This observation can be naturally accounted for by noticing that the spin of the final black hole is determined by the intrinsic and orbital angular momentum of point particles at the ISCO~\cite{Buonanno:2007sv}. Thus, particles with spins aligned have a stronger impact in the final black hole spin.
For charged particles a similar reasoning applies; namely, since the ISCO of charged particles moves inwards (and the ISCO frequency increases)
when charge is added to black holes, quasi-circular inspirals of equal-charge binaries would give rise to larger energy fluxes than
opposite-charged ones. Such scenarios would then provide ideal prospects for maximizing the gravitational energy output from the system.

\begin{acknowledgments}
We thank J.C.~Degollado for helpful discussions.
M.Z.\ is supported by NSF grants OCI-0832606, PHY-0969855, AST-1028087, and PHY-1229173.
V.C.\ acknowledges financial support provided under the European Union's FP7 ERC Starting Grant ``The dynamics of black holes:
testing the limits of Einstein's theory'' grant agreement no.\ DyBHo--256667.
U.S.\ acknowledges support by
the FP7-PEOPLE-2011-CIG CBHEO Grant No. 293412,
the STFC Grant No. ST/I002006/1,
the XSEDE Grant No. PHY-090003 by the National Science Foundation,
the COSMOS supercomputer infrastructure, part of the DiRAC HPC
Facility funded by STFC and BIS, and
the Centro de Supercomputacion de Galicia (CESGA) under Grant No. ICTS-2013-249,
This research was supported in part by Perimeter Institute for Theoretical Physics. 
Research at Perimeter Institute is supported by the Government of Canada through 
Industry Canada and by the Province of Ontario through the Ministry of Economic Development 
$\&$ Innovation.
This work was supported by the NRHEP 295189 FP7-PEOPLE-2011-IRSES Grant, and by FCT-Portugal through projects
PTDC/FIS/116625/2010 and CERN/FP/123593/2011.
Computations were performed on the ``Baltasar Sete-Sois'' cluster at IST, the ``Blafis'' cluster at Universidade de Aveiro, the NICS Kraken Cluster,
the SDSC Trestles Cluster, Cambridge's COSMOS,
on the ``venus'' cluster at YITP, and CESGA's Finis Terrae.

\end{acknowledgments}

\bibliographystyle{myutphys}
\bibliography{num-rel}

\end{document}